\documentclass[12pt,manuscript]{aastex}
\newcommand{\be}{\begin{equation}}
\newcommand{\ee}{\end{equation}}
\newcommand{\bea}{\begin{eqnarray}}
\newcommand{\eea}{\end{eqnarray}}
\shorttitle{Origin of Hot Jupiters}
\shortauthors{C. Beaug\'e and D. Nesvorn\'y} 
\begin{document}
\title{Multiple-Planet Scattering and the Origin of Hot Jupiters}
\author{C. Beaug\'e$^{1,2}$ and D. Nesvorn\'y$^2$}
\affil{(1) Observatorio Astron\'omico, Universidad Nacional de C\'ordoba, \\ 
Laprida 854, X5000BGR C\'ordoba, Argentina}
\affil{(2) Department of Space Studies, Southwest Research Institute, \\ 
1050 Walnut St., Suite 300, Boulder, CO 80302, USA}

\begin{abstract}

Doppler and transit observations of exoplanets show a pile-up of Jupiter-size
planets in orbits with a 3-day period. A fraction of these hot Jupiters have
retrograde orbits with respect to the parent star's rotation, as evidenced by
the measurements of the Rossiter-McLaughlin effect. To explain these
observations we performed a series of numerical integrations of planet
scattering followed by the tidal circularization and migration of planets that
evolved into highly eccentric orbits. We considered planetary systems having 3
and 4 planets initially (before the scattering phase), located at 1-15 AU, and
with masses between $0.5$ and $4$ times that of Jupiter. The simulations
included the tidal and relativistic effects, and precession due to stellar
oblateness.  Stellar and planetary tides were modeled in the approximation of
the time-lag equilibrium model that was modified for quasi-parabolic
orbits to include the main features of the dynamical tide model. We found that
the standard Kozai migration is an inefficient mechanism for the formation of
hot Jupiters, as the orbits typically acquire high eccentricities and high
inclinations due to close encounters and subsequent slow secular interactions,
rather than due to the sole effect of the Kozai resonance.

Our results show the formation of two distinct populations of hot Jupiters. The
inner population (Population I) of hot Jupiters with semimajor axis $a < 0.03$~AU 
formed in the systems where no planetary ejections occurred. This group
contained a significant fraction of highly inclined and retrograde orbits, with
distributions largely independent of the initial setup. However, our follow-up
integrations showed that this populations was transient with most planets
falling inside the Roche radius of the star in $<1$ Gyr. The outer population
of hot Jupiters (Population II) formed in systems where at least one planet was
ejected into interstellar space. This population survived the effects of tides
over $>1$ Gyr. The semimajor axis distribution of Population II fits nicely the
observed 3-day pile-up.

A comparison between our 3-planet and 4-planet runs shows that the formation of
hot Jupiters is more likely in systems with more initial planets. This appears
related to an increase in the chaoticity of the system and to a larger number
of close encounters. For example, the planetary systems with four initial
planets produce hot Jupiters twice as often as those with three
planets. Interestingly, the inclination distribution of Population II also
depends on the number of planets in the initial systems. While we found only a 
few hot Jupiters in retrograde orbits in the 3-planet case, the 4-planet case showed a
larger proportion (up to $10 \%$), and a wider spread in inclination values. As 
the later results roughly agrees with observations, this may suggest that the
planetary systems with observed hot Jupiters were originally rich in the number
of planets, some of which were ejected. In a broad perspective, our work
therefore hints on an unexpected link between the hot Jupiters and recently
discovered free floating planets.
\end{abstract}


\section{Introduction}

To date, the Rossiter-McLaughlin effect has been measured for 37 exoplanets
(Moutou 2011). Of these, 8 planets ($\sim20$ \%) are in retrograde orbits,
while in about half of the cases the planet orbit normal is probably aligned
with the stellar spin vector ($|\lambda| < 30^\circ$, where $\lambda$ is the
usual projected spin-orbit misalignment angle). Interestingly, the known
planets with $|\lambda| > 40^\circ$ have masses $M\lesssim 2$ M$_{\rm Jup}$,
where M$_{\rm Jup}$ is the mass of Jupiter, while planets with $M > 3$ 
M$_{\rm Jup}$ have $|\lambda| \gtrsim 40^\circ$ (Moutou 2011). If this trend holds
with the new data, it could provide an important hint on the dynamical origin
of the misaligned population. For example, as we will discuss in detail in
Section 3, planet scattering followed by tidal circularization and migration is
expected to produce such a trend because the less massive planets generally
evolve into more eccentric and inclined orbits than the more massive ones, and
are therefore more likely to show misalignment.

There also seems to be an indication that more circular orbits are accompanied
by small values of $|\lambda|$, while large values of $|\lambda|$ appear in
more eccentric cases (Schlaufman 2010). This can be interpreted as the evidence
for two distinct populations of close-in exoplanet systems. The former
population can be consistent with the smooth planetary migration in a gaseous disk
(e.g., Benitez-Llambay et al. 2011), while the later probably requires a more
complex orbital history.

To produce large values of $|\lambda|$, it is either necessary to tilt the spin
axis of the star so that it ends up being misaligned with the original
protoplanetary disk in which planets formed, or to tilt the planetary orbit. A
tilt of the star's axis could be produced by the cumulative effect of stellar
flares, late unisotropic (Bondi-Hoyle) accretion on the star (Throop \& Bally
2008, Moeckel \& Throop 2009), and due to the interaction between the stellar
magnetic field and the protoplanetary disk (Lai et al. 2011, and the references
therein). An orbital tilt can be produced by: (i) Planetary scattering followed
Kozai migration (e.g. Nagasawa et al. 2008); (ii) Kozai migration produced by a
planetary perturber in a distant and inclined orbit (Naoz et al. 2011); and
(iii) Secular migration in well-spaced, eccentric, and inclined planetary
systems (Wu \& Lithwick 2011).\footnote{Inclined stellar perturber could in
principle also trigger Kozai migration, but no distant star companions are
observed in systems with known hot Jupiters.}

Here we concentrate on the orbital tilt theories, because observational
evidence suggests that at least some planetary systems have large orbital
inclinations (e.g., planets $c$ and $d$ of $\nu$ Andromedae have mutual 
inclination $I\sim 30^\circ$; McArthur et al. 2010). In addition, the observed large
eccentricities of exoplanets can be best explain if the original packed
planetary systems underwent a dynamical instability followed by planet
scattering (e.g., Weidenschilling \& Marzari 1996, Rasio \& Ford 1996). As
planet scattering naturally leads to large orbital inclinations as well, the
orbital tilt theories are therefore a logical extension of the generally
accepted planet scattering model.

Theories (i) and (ii) listed above invoke the so-called Kozai mechanism, or
Lidov-Kozai resonance (Lidov 1961, Kozai 1962), to drive up the orbital
inclination $I$ towards values larger than ninety degrees. As the Lidov-Kozai
resonance appears in the secular dynamics of the three-body problem for mutual
inclinations $I \gtrsim 40^\circ$ (Libert \& Henrard 2007), a question arises
of how such a large inclination between planetary orbits can be achieved in the
first place. In addition, the retrograde orbits can be produced in the octupole
(and higher order) Kozai approximation only for a relatively small subset of
initial conditions. Naoz et al. (2011) did not address these issues in detail.

Wu \& Lithwick (2011) argued that the retrograde orbits can be achieved is
systems with a significant initial Angular Momentum Deficit (AMD) and well
separated planetary orbits. They showed that the smaller planets in these
systems can exchange angular momentum with the more massive planets and evolve
into highly-inclined, and possibly retrograde orbits, by the slow secular
interaction between orbits.  Therefore, this mechanism is similar to that
discussed in Naoz et al. (2011), but can potentially be more efficient than the
Kozai resonance in that it does not require finely-tuned initial conditions.
Still, it remains to be explain how the large AMD assumed by Wu \& Lithwick
(2011) arises in planetary systems as planets should form with a very low AMD.

The initial AMD can arise as a result of planet scattering, which brings us
back to the statistical study of Nagasawa et al. (2008). Somewhat ironically,
the initial setups of Naoz et al. (2011) and Wu \& Lithwick (2011) could
therefore be traced to planet scattering (unless alternative explanations are
offered for their initial conditions; e.g., Libert \& Tsiganis 2011ab).

Nagasawa et al. study was done before the first retrograde planets were
discovered so that, understandingly, not much effort was invested in their work
in a detail analysis of the mechanism that produced retrograde planets. This is
where the works of Naoz et al. (2011) and Wu \& Lithwick (2011) come to be
handy. As we discuss in detail in section 3, the Kozai mechanism alone appears
to be inefficient in producing the retrograde hot Jupiters. Instead, we find that
the orbital tilt is usually achieved through a combination of planet scattering,
and the secular chaos of Wu \& Lithwick (2011).

While many works considered planet scattering (e.g. Marzari \& Weidenschilling
2002, Chatterjee et al. 2008, Juric \& Tremaine 2008), Nagasawa et al. (2008)
pioneered the statistical studies of planet scattering {\it with tidal effects}.
The principal role of tides is the circularization of the planetary orbit while
approximately preserving the angular momentum. This effect helps to stabilize
the orbits of scattered planets reaching small pericenter distances ($q
\lesssim 0.05$ AU) and, consequently, leads to the formation of hot-Jupiters.

For the tidal effects, Nagasawa et al. (2008) adopted the dynamic tide model by
Ivanov \& Papaloizou (2004), which is applicable for fully convective planets
with near-parabolic orbits, but is not well suited for low-to-moderate
eccentricities. Conversely, Wu \& Lithwick (2011) used the equilibrium tide
model (e.g. Hut 1981, Mardling 2007) which is strictly valid only for
low-to-moderate eccentricities. Since the real evolution of planets likely
spans the whole range from near-parabolic to near-circular orbits, it is not
clear whether any of the two approximations mentioned above is adequate. We
explain how we deal with this problem in Section 2.

For the work described in this paper we assumed a physical scenario similar to
Nagasawa et al. (2008) starting with multiple-planetary systems in unstable
orbits and following their evolution through the stage of close encounters and 
planet scattering. Our main simulations employed an N-body code, incorporating 
the relativistic effects, stellar
oblateness and tidal precession. In Section 3 we describe our N-body code and
the results of scattering experiments with 3-planet systems. An extension to 4
planets is reported, fittingly, in Section 4. The long-term tidal evolution of
hot Jupiters is discussed in Section 5.

\section{Tidal Model}

\subsection{Equilibrium vs. Dynamical Tides}

Current tidal models were constructed for two limit cases. If the orbital
separation between the interacting bodies is roughly constant due to low
orbital eccentricity, the tides vary slowly and generate an equilibrium figure
in the extended bodies. Viscosity causes this tidal bulge to deviate from the
instantaneous equipotential shape which, in turn, leads to an angular momentum
exchange between the orbital and rotational motions. The dynamical evolution of
the bodies in this approximation is described by the so-called {\it equilibrium
  tide} model, originally developed by Darwin (1879). In its simplest version
(Mignard 1979, 1980), it is assumed that the equilibrium shape of each body at
time $t$ is defined by the equipotential surface at time $t + \delta t$. The
time lag $\delta t$ can be either positive or negative depending on whether the
rotational period is larger or smaller than the orbital period.

The opposite limit case occurs when $e \sim 1$. In this case the tidal
distortion is generated only at the pericenter and is negligible during the
rest of the orbit. Consequently, the bodies can no longer achieve equilibrium
figures. Instead, they undergo forced oscillations, the most important being
f-mode (or surface gravity) waves. The subsequent effects on the rotational and
orbital motion are described by the {\it dynamical tide} model (e.g. Lai 1997,
Ivanov \& Papaloizou 2004, 2007, 2011). Dynamical tides are much more complex
than their equilibrium counterparts and their effect on the orbital and
rotational evolution of the participating bodies is not so well understood.
For example, up to date only planar or polar orbits have been studied, and
there is no model for planets with arbitrary inclinations.

Unfortunately, the dynamical evolution that leads to formation of hot Jupiters
covers both limit cases discussed above. Initially, the planet has an almost
parabolic orbit caused by scattering and slow secular evolutions. Thus, the
tidal interaction with the star should be treated in the frame of the dynamical
tide model. As the orbit decays and circularizes, the system approaches the
equilibrium tide regime, and the dynamical tide model ceases to be valid.

Since the two models are based on a consideration of two completely different
physical phenomena (equipotential figures vs. damped forced oscillations), it
may be difficult to construct an unified physical tidal model that would be adequate 
for the entire orbital evolution of hot Jupiter. To deal with this problem we 
empirically modified the equations of the equilibrium tide model to mimic the 
effects of the dynamical tide model when the parabolic limit is approached.

\subsection{Equilibrium Tidal Model}

We use the equilibrium tidal equations derived in Correia et al. (2011). These
equations are based on the linear Mignard model, and contain explicit
expressions for the variations of the mutual inclination, argument of the
pericenter and obliquities, all in a consistent manner. They also include
additional perturbations from gravitational interactions with other planets,
stellar oblateness and general relativity in the post-Newtonian approximation.

Tidal precession is usually neglected in tidal equations, since the
precessional period is much shorter than the timescale associated to
circularization and orbital decay. However, since it is expected that
Lidov-Kozai resonance may play an important role in the orbital evolution of
hot Jupiters, we retain these secular terms in our model.

We introduce two important changes with respect to Correia et
al. (2011). First, instead of limiting the gravitational interactions between
the planets to the quadrupole secular approximation, we extend it to the
octupole level. Second, instead of adopting Jacobi coordinates, we use
Poincar\'e astrocentric coordinates.

We assume three extended bodies: $m_0$ (star), $m_1$ (inner planet) and $m_2$
(outer planet), with the planets having orbital elements $a_i$, $e_i$, $I_i$
and $\omega_i$, where $\omega_i$ are arguments of pericenters. Tidal effects
will only be felt by $m_0$ and $m_1$. The outer body is assumed to be too far
from the central star to generate or receive tidal distortions, but will
interact gravitationally with $m_1$.

We use the following variables for the orbital motion:
\begin{eqnarray}
\label{eq1}
{\bf G_i}  &=& \beta_i \sqrt{\mu_i a_i (1-e_1^2) } {\bf {\hat k}_i}  \\
{\bf e_i}  &=& \frac{{\bf {\dot r}_i} \times {\bf G_i}}{\beta_i \mu_i} - 
\frac{{\bf r_i}}{r_i} \nonumber
\end{eqnarray}
where $\beta_i$ and $\mu_i$ are the Poincar\'e mass factors (e.g. Laskar \&
Robutel 1995), ${\bf G_i}$ are the orbital angular momentum vectors for each
mass and ${\bf e_i}$ the Lenz vectors. The unit vector ${\bf {\hat k}_i}$ is
perpendicular to the orbital plane (the reference plane is arbitrary), while
the direction of the Lenz vector points towards the argument of pericenter.

Additionally, for the tidally interacting bodies, we also define the rotational
frequencies by $\Omega_0$ and $\Omega_1$. The equations of motion for the
objects' spins will be written in terms of the rotational angular momenta:
\begin{equation}
\label{eq2}
{\bf L_i}  = C_i \Omega_i {\bf {\hat s}_i}
\end{equation}
where $C_i$ are the principal moments of inertia and ${\bf {\hat s}_i}$ the
spin axis referred to the same reference plane. It is assumed that the spin
vectors always coincide with the orientation of the principal moments of
inertia.

Correia et al. (2011) constructed the variational equations for the complete
set of variables $({\bf G_i},{\bf e_i},{\bf L_i})$ that include all the
above-mentioned perturbations, from secular terms of the mutual gravitational
interactions between the planets to tidal effects, GR and stellar
oblateness. At any given instant, the eccentricities can be obtained from the
modulus of the Lenz vector, the semimajor axis from $|{\bf G_i}|$ and the
rotational frequencies from $|{\bf L_i}|$. The mutual inclination of both
orbits can be calculated from:
\begin{equation}
\label{eq3}
\cos{I_{\rm mut}} = {\bf {\hat k}_1} \cdot {\bf {\hat k}_2}.
\end{equation}
Finally, the obliquities can be determined from:
\begin{equation}
\label{eq4}
\cos{\theta_i} = {\bf {\hat s}_i} \cdot {\bf {\hat k}_1}  
\hspace*{0.5cm} ; \hspace*{0.5cm}
\cos{\varepsilon_i} = {\bf {\hat s}_i} \cdot {\bf {\hat k}_2} ,
\end{equation}
where $\theta_i$ is the obliquity of $m_i$ with respect to the orbital plane of
the inner body, while $\varepsilon_i$ are the obliquities with respect to the
orbital plane of the outer companion.  We will be most interested in variables
$\theta_0$ and $\theta_1$.

The orbital/rotational equations of motion yield a complete description of the
tidal and gravitational evolution of the system, valid as long as no resonances
or close encounters between the planets are expected. It then constitutes a
semi-analytical model that requires much less CPU time than a direct N-body 
integration.

\begin{figure}[t]
\includegraphics[width=0.45\textwidth,clip=true]{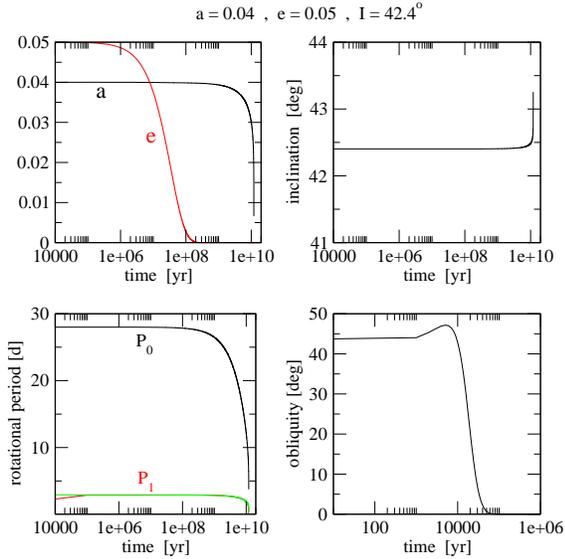}
\caption{Tidal evolution of a Jovian mass planet with initial semimajor axis 
$a=0.04$ AU, $e=0.05$ and $I=45^\circ$ with respect to the 
stellar equator. $P_0$ is the stellar rotation period, while $P_1$ is the planet's 
rotation period. The green line in the bottom-left panel shows the planet's 
orbital period. The planet's obliquity shown in the bottom-right panel is 
measured with respect to the planetary orbital plane. The initial spin axis 
of the planet was assumed to coincide with that of the star.}
\label{fig1}
\end{figure}

\begin{figure}[t]
\includegraphics[width=0.45\textwidth,clip=true]{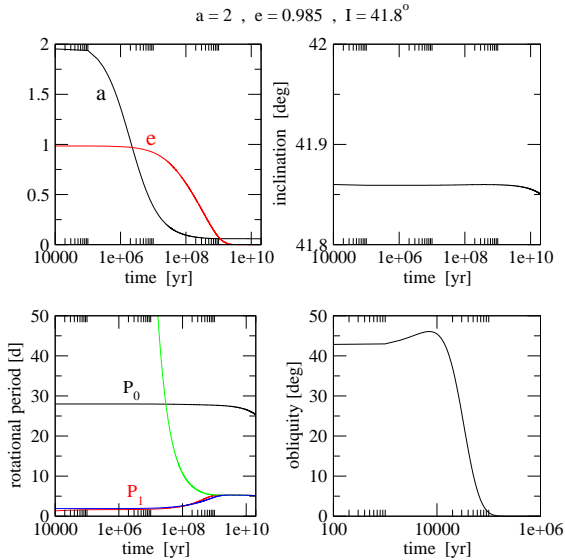}
\caption{Same as Figure \ref{fig1}, but for an initial orbit with $e\sim1$. 
The blue line in the bottom left panel shows $2 \pi /n_q$, where 
$n_q=\sqrt{{\cal G}/q^3}$.}
\label{fig2}
\end{figure}

Figures \ref{fig1} and \ref{fig2} show two examples, both with $m_0= M_\odot$
and $m_1=m_1={\rm M}_{\rm Jup}$. For these illustrations, we used $Q'_*=10^7$ 
and $Q'_p=10^5$ for the star and planet tides, respectively. The quantity
$Q'=Q/k_2$ is the so-called modified tidal parameter, where $k_2$ is the tidal
Love number. The value for the stellar parameter $Q'_*$ was taken from
Benitez-Llambay et al. (2011) as the value that best fits the semimajor axis
distribution of short-period planets. The value for $Q'_p$ follows from the recent
determination for Jupiter (e.g. Lainey et al. 2009; but see Section 2.4 where
we find $Q'_p=(1$-$5)\times10^6$ for hot Jupiters).

In Fig. \ref{fig1}, the initial semimajor axes and eccentricity is similar to
that of the known hot Jupiters ($a=0.04$ AU and $e=0.05$). In Fig. \ref{fig2},
we assumed an initially quasi-parabolic orbit with larger semimajor axis ($a=2$
AU and $e=0.985$) similar to those obtained from scattering experiments
(e.g. Nagasawa et al. 2008). In the latter case, $q=1(1-e)=0.03$~AU, thus
placing the planet's pericenter within the region affected by tides. In both
cases $m_2$ was placed in a distant circular orbit at $a_2=100$ AU to make its
gravitational perturbations on the inner planet insignificant. 
The mutual inclination was chosen $I=45^\circ$, and both initial spin vectors 
were set to be perpendicular to the reference plane. Finally, the initial rotational 
periods were $P_0 = 2 \pi / \Omega_0 = 28$ days and $P_1 = 2 \pi / \Omega_1 = 0.4$ 
days.

Both simulations show practically no change in the orbital inclination due to
tides, except during the final stages just before the planet is engulfed by the
star. For quasi-circular orbits, the orbital decay time $\tau_a$ is longer
than the circularization timescale $\tau_e$, implying that the planet is
circularized before becoming a hot Jupiters (Fig. \ref{fig1}). The opposite
occurs for the initially quasi-parabolic orbits, where $\tau_e > \tau_a$
(Fig. \ref{fig2}). This latter case is consistent with the results of Ivanov \&
Papaloizou for dynamical tides (see also Nagasawa et al. 2008). However, even
in this case, once the eccentricity decreases to small values, the relation
switches back and the final stage of the orbital evolution occurs as in Figure
\ref{fig1}.

Another difference in the high eccentricity case is that the synchronization of
the planetary spin occurs with respect to the orbital frequency at pericenter
and not with respect to $n=\sqrt{{\cal G}/a^3}$. Once again, this is in good
agreement with the predictions of Ivanov \& Papaloizou for the dynamical tide
model. Thus, it appears that several of the orbital evolutionary properties of
the dynamical tide model can be fairly and qualitatively reproduced with the
equilibrium tide approximation.

Finally, we discuss the behavior of the planetary obliquity
$\theta_1$. Planetary tides cause a relatively rapid alignment of the planet's
spin axis with respect to the orbital plane normal, thus leading to $\theta_1
\sim 0$. At least for the simulations discussed in this paper, we found no
cases of trapping in other Cassini states. This occurs even for the retrograde
orbits, where the initially obliquity was larger than $90^\circ$. However, the
time scale of obliquity evolution is of the order of $10^5$ years. This is much
larger than the synchronization timescale ($\sim 10^4$ yr), but much smaller
than the orbital decay times. Thus, in the absence of additional forces,
it is expected that $\theta_1 \sim 0$ during most of the planet's evolution. 

\begin{figure}[t]
\includegraphics[width=0.45\textwidth,clip=true]{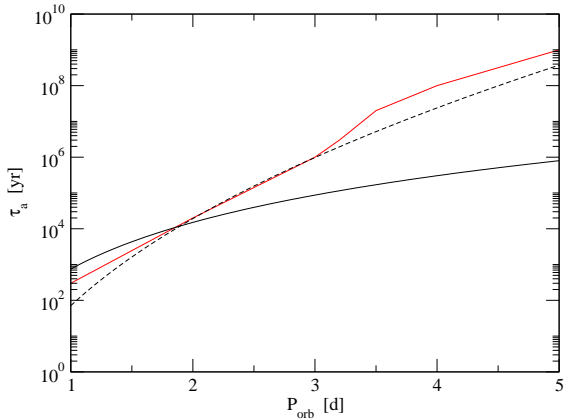}
\caption{Orbital decay timescale as a function of the orbital period after 
circularization, for a Jovian mass planet in an initial quasi-parabolic orbit. 
Black curve is the prediction of the equilibrium tide model, while the red curve 
is the numerical estimate from the dynamical model from Ivanov \& Papaloizou 
(2011). The dashed curve was obtained from the equilibrium tide model by 
modifying the tidal parameters according to equation (\ref{eq5}).}
\label{fig3}
\end{figure}

\subsection{Correction Terms in the Tidal Model}

One of the most important differences between the equilibrium and dynamical
tide models is their implication for the orbital decay and orbital
circularization e-folding times ($\tau_a$ and $\tau_e$, respectively). Figure
\ref{fig3} shows the value of $\tau_a$ obtained from the equilibrium tide model
(same tidal parameters as used in the previous figures) and from the dynamical
model of Ivanov \& Papaloizou (2011) (see their Figure 5). The decay times are
plotted as a function of $P_{\rm orb}$ defined as the planet's orbital period
after circularization. The dynamical model predicts shorter decay times for
$P_{\rm orb} \le 2$ days and suggests much longer timescales for longer
periods.

A way to reproduce, at least qualitatively, the results of Ivanov \& Papaloizou
(2011) with the equilibrium tide model is to modify the values of the tidal
parameters according to the following empirical recipe: 
\begin{equation}
\label{eq5}
Q'_i \longrightarrow Q'_i \; 10^{\beta}  \hspace*{0.3cm} {\rm with} \hspace*{0.3cm}
\beta = 2 e^2 (a-3/2) .
\end{equation}
The factor $e^2$ guarantees that this change is only significant for highly
eccentric orbits, since for the lower eccentricity values the equilibrium tide
model works well. Moreover, we implemented the correction term only for orbits 
with $a>1$ AU. This is because the dynamical tide is expected to be relevant 
in our simulations only for large semimajor values.

The value of $\tau_a$ determined with this new tidal parameter is also plotted
in Figure \ref{fig3}. The overall agreement is satisfactory. The
correction in Eq. \ref{eq5} also affects the circularization timescale
$\tau_e$, yielding values very close to those predicted by Ivanov \& Papaloizou
(2011).

\subsection{Constraint on $Q'_p$}

\begin{figure}[tb!]
\includegraphics[width=0.45\textwidth,clip=true]{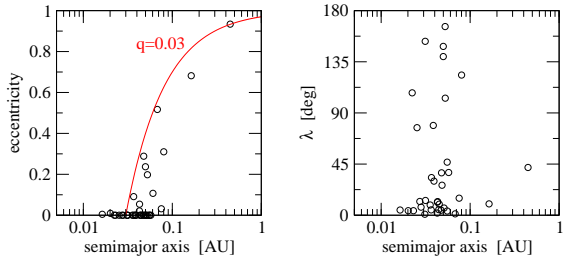}
\caption{Distribution of known exoplanets with measured $\lambda$. Left plot 
shows eccentricity as a function of the semimajor axis. The red line denotes constant 
pericentric distance $q=0.03$~AU. Right plot shows $\lambda$ as function of 
the semimajor axis.}
\label{fig4}
\end{figure}

\begin{figure}[tb!]
\includegraphics[width=0.45\textwidth,clip=true]{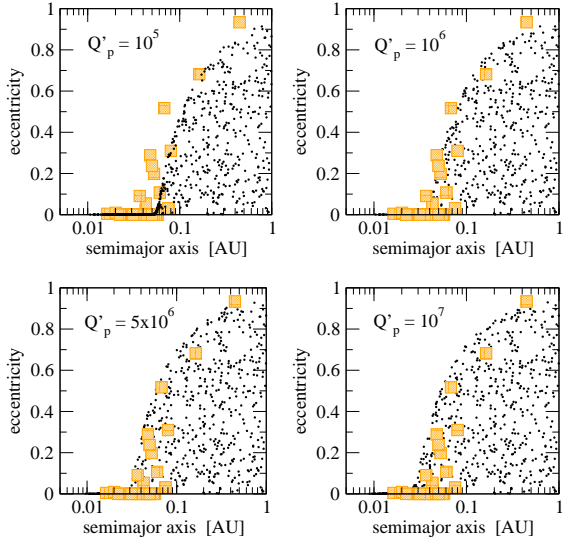}
\caption{Eccentricity-semimajor axis distribution expected from tidal 
models with four different values of $Q'_p$. 
Orange squares show the distribution of real exoplanets with measured 
$\lambda$ (data from {\rm www.exoplanet.eu} webpage).}
\label{fig5}
\end{figure}

\begin{figure}[tb!]
\includegraphics[width=0.45\textwidth,clip=true]{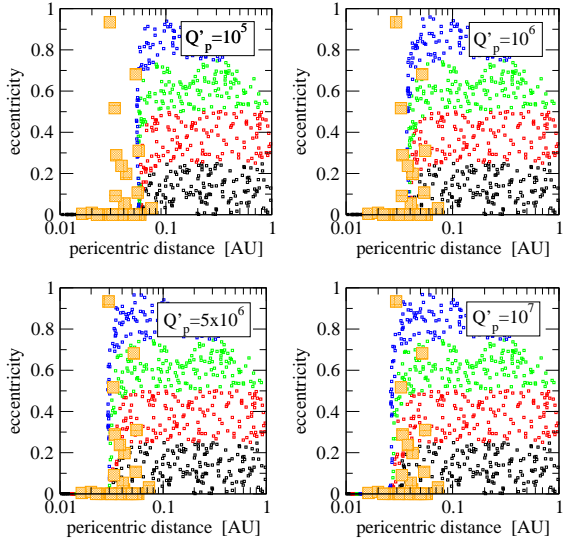}
\caption{As Figure \ref{fig5}, but this time showing the 
final eccentricity in terms of the final pericentric distance. The color code 
identifies the initial eccentricity of each planet: black for those initial 
conditions with $e<0.25$, red for $0.25 < e < 0.5$, green for $0.5 < e < 0.75$ 
and blue for $e > 0.75$. Th orbits of real exoplanets are labeled by 
orange squares.}
\label{fig6}
\end{figure}

Figure \ref{fig4} shows the distribution of known exoplanets with measured
$\lambda$. Interestingly, the eccentricities show a marked correlation with the
semimajor axis, roughly along a line with $q=0.03$ AU.  This could potentially
be used to constraint the value of $Q'_p$.

To do that, we followed the tidal evolution of Jovian-mass planets. Their
initial semimajor axis was taken from a uniform distribution in $\log(a)$ from
$0.01$ AU to $4$ AU. The eccentricities were taken from a uniformly random
distribution of $e$ between zero and one. A total of $10^4$ initial orbits
were generated. Each orbit was evolved for $1$ Gyr using our hybrid model. Results are
shown in Figure \ref{fig5} for four different values of the planetary tidal
parameter $Q'_p$, ranging from $10^5$ (top left) to $10^7$ (bottom right).

For $Q'_p = 10^5$, the envelope of final orbits does not seem to reproduce the
observed eccentricity distribution. The tidal effects are apparently too
efficient, and all orbits with final semimajor axes below $0.06$ AU become
circularized. Better results are obtained with $Q'_p = 10^6$ and $Q'_p = 5
\times 10^6$, although larger values seem to be too inefficient.

If the real exoplanets are the outcome of scattering events and tidal capture
from quasi-parabolic orbits, it is expected that they underwent a phase when
their orbital eccentricities were were very high. 
Figure \ref{fig6} factors this assumption in the results of our
tests. Once again the standard equilibrium tide model with $Q'_p=10^5$ does not
yield a good agreement, but now it is clear that the modified model with
$Q'_p=10^7$ is also not adequate, since practically all the real planets fall
into the region with relatively small initial eccentricities.

From these tests it appears that the eccentricity-semimajor axis
distribution of the real exoplanets can be reproduced by planetary tidal
parameters of the order of $Q'_p \sim (1$-$5) \times 10^6$.  Similar values have
been proposed by other authors. Using similar analysis, Jackson et al. (2008)
found that the distribution of all close-in planets shows good agreement
assuming $Q'_p \simeq 3 \times 10^6$, while a value of $Q'_p \simeq 2.2 \times
10^6$ has recently been proposed to explain the current orbital characteristics
of CoRoT-20b (Deleuil et al. 2011).

For the present work we mainly used $Q'_p = 5 \times 10^6$, and tested a range of $Q'_p$ 
values in Section 3, and $Q'_*=10^7$. As long as the planet has a non-negligible 
orbital eccentricity, the stellar tidal parameter $Q'_*$ is of secondary importance 
because the orbital dynamics is mainly controlled by the planetary tide. However, 
the stellar tide is important once the planet reaches a circular orbit.

\section{N-Body Simulations of 3-Planet Scattering}

To study the formation of hot planets, we must first follow the orbital
evolution of a planetary system through the instability phase, when planets
scatter off of each other. The code must also be able to track the evolution in
the late stage when hot Jupiters become circularized and migrate by tides, as
discussed in Section 2. Here we describe our N-body integrator that is able to
handle close encounters between planets, near-parabolic orbits that may result
from encounters, and tides. To start with, we will apply this code to 3-planet
systems, a case similar to that studied by Nagasawa et al. (2008).

\subsection{Explicit Expressions for the Forces}

In an astrocentric reference frame, the differential equation affecting the
position vector ${\bf r}$ of the planet is:
\begin{equation}
\label{eq9}
{\ddot {\bf r}} = {\bf f}_0 + {\bf f}_{TP} + {\bf f}_{TD} + {\bf f}_{GR} + {\bf f}_{SO} \ ,
\end{equation}
where
\begin{equation}
\label{eq10}
{\bf f}_0 = - \frac{\mu}{r^3} {\bf r} 
\end{equation}
is the gravitational acceleration from the central star, and $\mu = {\cal G}
(m_0 + m_1)$. The tidal distortion that generates the apsidal precession (see
Hut 1981) is given by
\begin{equation}
\label{eq12}
{\bf f}_{TP} = -3 \frac{\mu}{r^8}\biggl[ {k_2}_0 \biggl( \frac{m_1}{m_0} 
              \biggr) R_0^5 + {k_2}_1 \biggl( \frac{m_0}{m_1}  
              \biggr) R_1^5 \biggr] {\bf r}\ ,
\end{equation}
while the corresponding tidal dissipation term is
\begin{eqnarray}
\label{eq13}
{\bf f}_{TD} &=& -3 \frac{\mu}{r^{10}} \sum_{i=0,1} \frac{m_{1-i}}{m_i} {k_2}_i 
                 \Delta_i R_i^5 \nonumber \\
            & & \biggl( 2 {\bf r} ({\bf r} \cdot {\dot {\bf r}}) + r^2 
 ({\bf r} \times {\bf \Omega}_i + {\dot {\bf r}}) \biggr) .
\end{eqnarray}
Here ${\bf \Omega}_i$ are the spin vectors of both bodies with respect to the
reference frame that does not have to coincide with the Laplace plane. The
factor ${k_2}_i \Delta_i$ is related to the tidal parameter through:
\begin{equation}
\label{eq14}
{k_2}_i \Delta_i = \frac{3}{2Q'_i n}
\end{equation}
where $n=\sqrt{\mu/a^3}$ is the mean motion of the planet. We used the hybrid
equilibrium tidal model discussed in the previous section with $Q'_*=10^7$ and
$Q'_p= 5 \times 10^6$.

Finally, 
\begin{equation}
\label{eq11}
{\bf f}_{GR} = -\frac{\mu^2 a (1 - e^2)}{c^2 r^5} {\bf r}\ ,
\end{equation}
where $c$ is the speed of light, is the post-Newtonian radial term that
approximates the General Relativity effects, and
\begin{equation}
\label{eq14bis}
{\bf f}_{SO} = \nabla \biggl( \mu J_2 \frac{R_0^2}{r^3} \biggl( \frac{3}{2} 
\sin{\delta} - \frac{1}{2} \biggr) \biggr)
\end{equation}
is the acceleration from stellar oblateness (e.g. Beutler 2005), where $R_0$ is
the stellar radius, $\delta$ the declination angle of the orbit with respect to
the stellar equator, and $J_2$ the quadrupole coefficient.

From the complete expression for the acceleration we can calculate its effect
on the spins, assuming a conservation of the complete (orbital + rotational)
angular momentum (see Correia et al. 2011 for more details).  We disregard any
variation of the stellar rotational frequency due to momentum loss driven by 
stellar winds (e.g., Skumanich 1972), and assume that it is only affected by tides.

\subsection{The Code}

Our original plan was to implement these forces into {\it SyMBA} (Duncan et
al. 1998), which is an efficient symplectic $N$-body code that is capable of
tracking close encounters between massive bodies. As {\it Mercury} (Chambers
1999), {\it SyMBA} uses the Poincar\'e variables to be able to handle
encounters. The symplectic algorithms written in Poincar\'e variables, however,
have troubles in following orbits with very high eccentricities. This is a
problem, because the orbits of hot Jupiters are expected to have $e\sim1$
before they can become circularized by tides.

To solve this problem, Levison \& Duncan (2000) proposed a hybrid integration
scheme in which the outer part of the eccentric planetary orbit is integrated
with the usual {\it SyMBA} algorithm. The code then symplectically switches to
the Bulirsch-Stoer (BS) algorithm to follow the planet's evolution near the
inner part of its orbit. The switch radius is set to a fixed apocentric
distance, usually of order of 0.1 AU.

We tested the hybrid algorithm in the extreme case when $e\sim1$ and found,
perhaps not surprisingly, that the time-step needs to be set to a very small
fraction of the orbital period. This slows down the symplectic algorithm so
much that the Bulirsch-Stoer (BS) integrator, if used to follow the full
evolution of the system (i.e., also outside the switch radius), is actually
faster and more precise. For this reason, we constructed a new $N$-body code
based on the BS method and used it in this study.

The code follows the interaction of $N$ massive planets orbiting a central star of
mass $m_0$. It tracks both the orbital and spin dynamics according to the
equations given in Section 3.1.  As in Nagasawa et al. (2008), the integrations
were stopped when reaching $10^8$ years, or if:

\begin{itemize}

\item A hot Jupiter formed with $e < 0.01$ and stable orbit.

\item One planet was ejected and the other two remained in stable orbits. In
  this case, the evolution of the system was continued using the secular model
  described in Section 2.
The use of the semi-analytical model helped to speed up the simulation, and
proved adequate for two-planet systems where no additional close encounters
between planets occurred.

\item Two planets were ejected and the pericentric distance of the survivor was
  larger than $0.1$ AU.

\end{itemize}

To determine whether a surviving pair of planets attained stable orbits we used
the Hill criterion of Marchal \& Bozis (1982) (see also Gladman 1993).

\subsection{Initial Conditions}

Planetary migration produced by planet--gas-disk interactions is an important
evolutionary process during the early history of planetary systems. As we do
not model these early stages here, we will need, in an uncertain leap of faith,
to adopt some initial conditions for our simulations. These condition should be
at least broadly consistent with the state of the planetary systems just after
the gas disk dispersal.

While classical hydrodynamical simulations of disk-planet interactions and
resonance trapping have focused on two-planet systems (e.g. Snellgrove et
al. 2001, Kley 2003), recent studied have been extended to multiple planetary
systems (e.g. Morbidelli et al. 2007, Libert \& Tsiganis 2011a, 2011b). It
appears that multiple-resonance trapping is a common outcome, although not all
the resonant configurations are long-term stable.  Thommes et al. (2008) also
suggested that stable configurations within the gas disk may become unstable
after disk dispersal and subsequent planetary scattering may occur.

We used several different mass ratios between planets. The masses in the units
of $m_{\rm Jup}$ were:
\begin{eqnarray}
\label{eq16}
m_1 &=& 1  \\ 
m_{(i+1)} &=& H \, m_i  \hspace*{0.5cm} (i=2,3) \nonumber
\end{eqnarray}
where $H=0.5$, $1$ or $2$. After the mass values were specified (using a random
generator for the $H$ values), we shuffled the radial order of the planets,
such that the body identified as $m_1$ was not necessarily the one closest to
the star.

The semimajor axis of the inner planet was chosen randomly between $1$ and $5$
AU. The other planets were placed in successive mean-motion resonances. We used
the $2/1$, $3/2$ or $4/3$ resonant ratios for different planet pairs, with the
specific choice depending on the individual masses of the two planets. To
select the resonant ratio that should apply to a specific mass ratio, we
adopted the results of Pierens \& Nelson (2008), who studied the resonant
capture with a hydrocode. Using the same disk parameters as Pierens \& Nelson
(2008), we also performed additional hydrocode simulations with {\it FARGO} (Masset
2000) to confirm and extend these results to multi-planet systems.

Orbital eccentricities were chosen randomly between zero and $0.1$ and
inclinations between zero and $1$ degree. Although this seems arbitrary, we
performed tests with different distributions (e.g. lower eccentricities 
for more massive planets) and found that the results were largely independent
of these assumptions. The
angular variables were randomly changed from their resonant values to mimic the
situation at the onset of instability. All initial conditions were consequently
dynamically unstable and none remained in the initial resonant configuration.

\subsection{Results}

\begin{figure}[tb!]
\includegraphics[width=0.45\textwidth,clip=true]{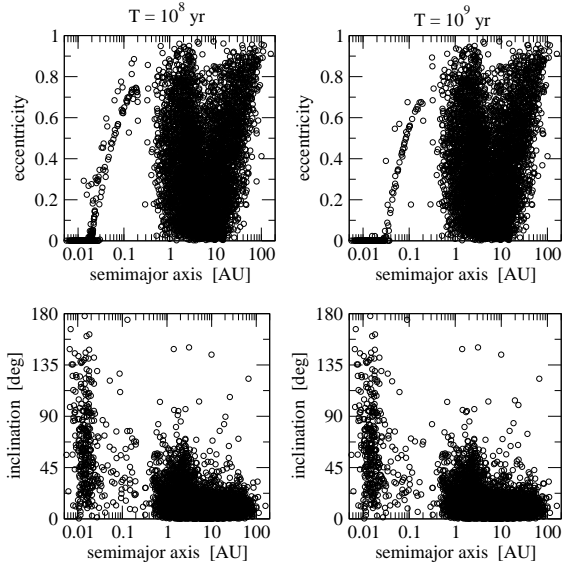}
\caption{Left: Results of our 2464 N-body integrations of the 3-planet systems 
with $Q'_*=10^7$ and $Q'_p=5 \times 10^6$. The total integration time was 
set to $T=10^8$ yrs. Right: Extension to $T=10^9$ yrs using the semi-analytical 
model. We stopped the simulation if hot Jupiter formed and acquired a nearly circular orbit 
(defined as $e<0.01$).}
\label{fig7}
\end{figure}

In total, we followed the evolution of 2464 initial systems, a number
sufficiently large for a detailed statistical analysis. Figure \ref{fig7} shows
the orbital distribution of all planetary systems at the end of simulations. A
total of 288 hot Jupiters formed, which is approximately $11 \%$ of the number
of initial systems. This fraction is about three times smaller than what
was reported in Nagasawa et al. (2008).
Most hot Jupiters acquired circular orbits due to tidal damping. Some hot
Jupiter, particularly those with $a\gtrsim 0.03$ AU, retained high
eccentricities.

\begin{figure}[tb!]
\includegraphics[width=0.45\textwidth,clip=true]{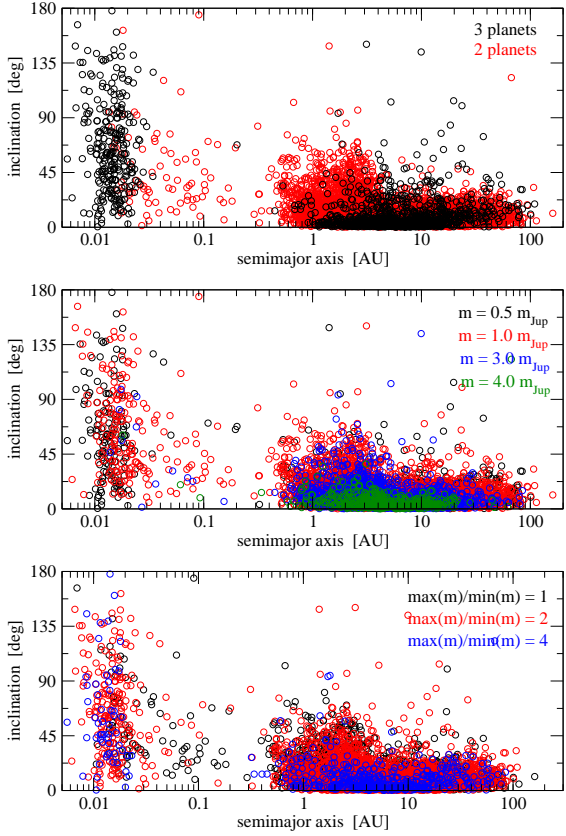}
\caption{Distribution of the final orbital inclination as function of the 
semimajor axis. Top: Black circles indicate the planetary systems where no 
ejections occurred, while red symbols show those in which one planet was ejected. 
Middle: Color indicates the mass of each planet (see inlaid color code caption for 
details). Bottom: Color indicates the planet mass ratio defined as the ratio between 
the largest and smallest planet masses in the original systems.}
\label{fig8}
\end{figure}

The left-hand panels in Figure \ref{fig7} show the results of the $N$-body simulation for a total
integration time of $T=10^8$ yrs. Since most of the known hot Jupiters
with measured $\lambda$ have stellar ages of the order of 1 Gyr (e.g. Triaud
2011), we extended the simulations to see how the population of hot Jupiter can
be modified by tidal effects over Gyr-long time scales.
First, we selected the planetary systems, where the $N$ body integrations
described above led to the formation of hot Jupiter. We disregarded the outer
planets in each system because their interaction with the hot Jupiter is
weak. We then used the semi-analytical code described in Section 2 to follow
the tidal evolution of hot Jupiters up to $T=1$ Gyr. The results are shown in
the right-hand plots of Figure \ref{fig7}. 

Note that we stopped the simulation if a hot Jupiter evolved into a nearly circular 
orbit (defined as $e<0.01$). The results shown in Figure \ref{fig7} therefore
include all hot Jupiters that formed in our simulations {\it at any time}.
As we will discuss in Section 3.6 and 4, many of these hot Jupiters evolve by 
tides and are dynamically short lived.

As for the eccentricity distribution, we find two types of final orbits. The
first type consists of planets that where tidally trapped early in the
simulations and had sufficient time to undergo tidal circularization. They show
up in Fig. \ref{fig7} as having nearly circular orbits. The second orbit type
shows moderate to large eccentricities (in some cases as high as $e\sim0.8$),
and a clear correlation between $a$ and $e$. Such a correlation is expected for a
population of planets evolving from quasi-parabolic orbits.

As for the inclination distribution, we note two distinct populations of
hot Jupiters that can be conveniently classified as having $a \leq 0.03$ AU
(Population I or Pop I) and $a > 0.03$ AU (Pop II). Practically all orbits with
$a > 0.03$ AU have $I<90^\circ$, while about $22 \%$ of those with $a < 0.03$
AU are retrograde.

A second difference between Pop I and Pop II is the behavior of planet's
obliquity. Practically all Pop-I planets have zero obliquities (relative to
planet's orbit), as expected from a sustained tidal evolution. In contrast,
many Pop-II planets retained relatively large values of $\theta_1$. These
planets continue to evolve by tides even $1$ Gyr after their parent system's
formation.

Next we studied the effect of planet masses and starting semimajor axes on the
results (Figure \ref{fig8}). We find that Population I mainly contains planets that
formed through scattering in systems in which no planet was ejected. The Pop-II
planets, on the other hand, formed in systems where one planet was ejected. We
discuss this interesting result in more detail in Section 3.5.

The middle frame in Figure \ref{fig8} shows the $a$ and $I$ distribution sorted
according to the planetary mass. This plot shows that the more massive planets
remain near their original orbits at $a>1$ AU. The hot Jupiters, on the other
hand, tend to be the least massive planets in the original systems. This is
easy to understand because the lighter planets are easier to scatter, and more
likely to become hot Jupiters. Although there seems to be no significant
difference in the inclination distribution between $m=0.5$ M$_{\rm Jup}$ and $m=1$ 
M$_{\rm Jup}$, the inclinations attained by the more massive planets are noticeably
smaller.

The bottom panel in Figure \ref{fig8} separates the final planets according to
the planet mass ratio in the initial systems. The results indicate that the Pop-I
planets tend to form in planetary systems with a large mass ratio, while the
outer Pop-II planets primarily evolve from the systems with planets more similar in
mass.

\begin{figure}[tb!]
\includegraphics[width=0.47\textwidth,clip=true]{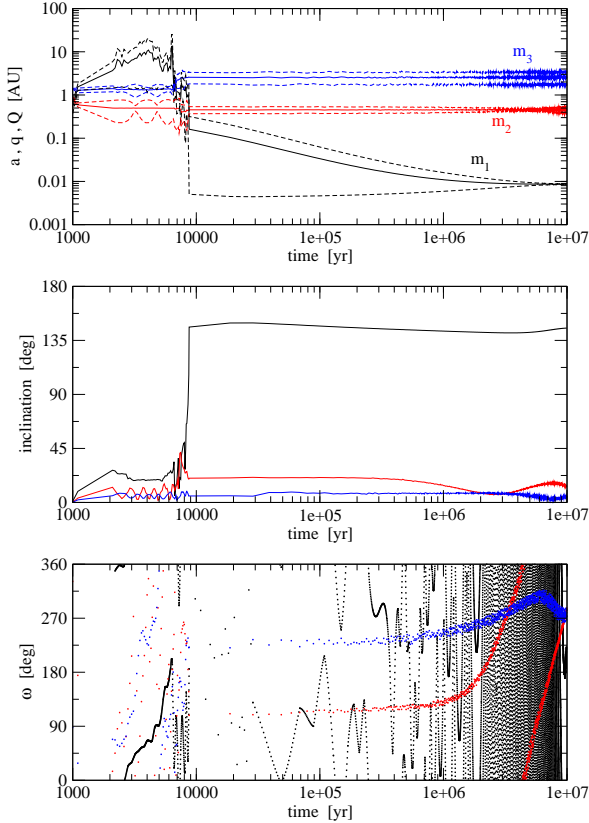}
\caption{Orbital evolution of a 3-planet system leading to the formation 
of hot Jupiter on a retrograde orbit. Top: semimajor axis and pericentric/apocentric 
distance. Middle: orbital inclinations. Bottom: Pericenter arguments. Planetary 
masses were chosen equal to $m_1=1$~M$_{\rm Jup}$, $m_2=2$ M$_{\rm Jup}$ and 
$m_3=4$ M$_{\rm Jup}$.}
\label{fig9}
\end{figure}

\begin{figure}[tb!]
\includegraphics[width=0.47\textwidth,clip=true]{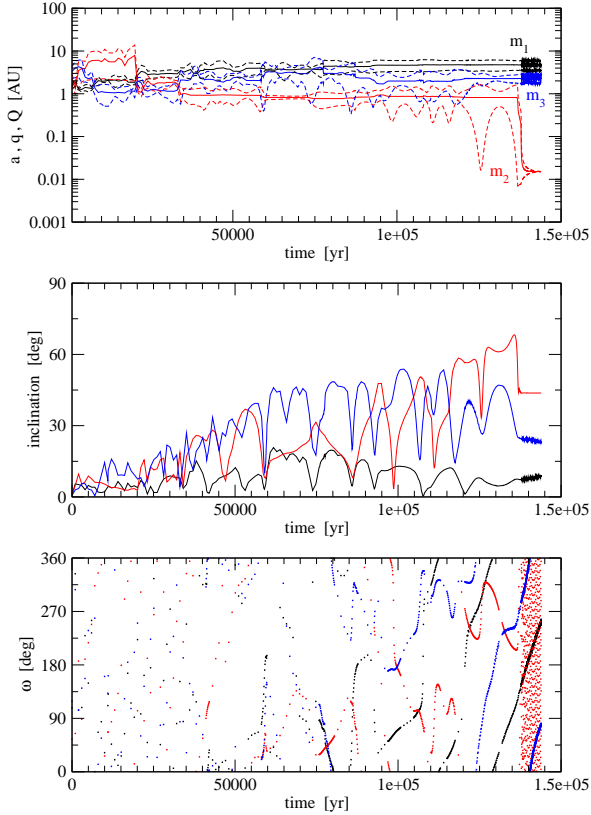}
\caption{As Figure \ref{fig9}, but for a system that produced hot Jupiter on
  a prograde orbit.  Note that the scale of the middle plot has changed with
  respect to Figure \ref{fig9}. Also, in this case, planetary masses were
  $m_1=1$ M$_{\rm Jup}$, $m_2=0.5$ M$_{\rm Jup}$, and $m_3=0.5$ M$_{\rm Jup}$.}
\label{fig10}
\end{figure}

\begin{figure}[tb!]
\includegraphics[width=0.47\textwidth,clip=true]{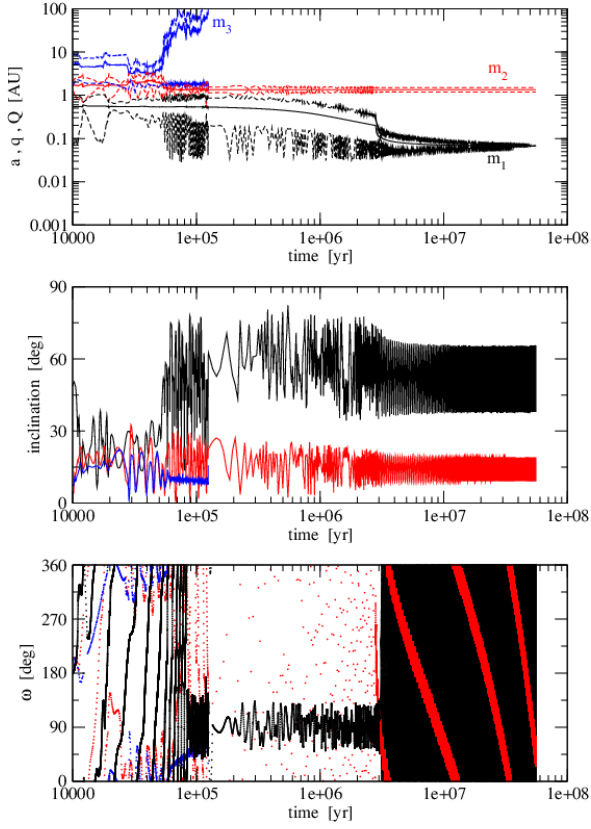}
\caption{Example of a rare planetary system showing clear effects of the
  Lidov-Kozai resonance.  Pop-II hot Jupiter formed in this
  simulation. Planetary masses were $m_1=1$~M$_{\rm Jup}$, $m_2=2$ M$_{\rm
    Jup}$, and $m_3=1$ M$_{\rm Jup}$.}
\label{fig11}
\end{figure}

\subsection{Orbit evolution}

To understand the dynamical mechanism responsible for the formation of hot
Jupiters, we now discuss the dynamical evolution of individual systems. We
illustrate things on three different planetary systems.

Figure \ref{fig9} shows the first case. In this case, none of the planets
escapes, which is characteristic for the systems in which Pop-I hot Jupiters
formed in our simulations. The initially inner planet ($m_1$ shown in black) is
quickly scattered to an exterior orbit. At $t\simeq 7000$ yrs, it suffers a
sequence of close encounters with the other two (more-massive) planets, a
period that lasts several thousand years.  In consequence, the lighter planet
is scattered into an inner orbit, its eccentricity raises to $\sim1$, and orbital
inclination very rapidly reaches values larger than $90^\circ$.

In the process, the pericentric distance drops and the planet decouples from
other planets in the system to evolve solely by tides over the rest of its
lifetime. Both the semimajor axis and eccentricity decrease due to tides while
maintaining similar values of the pericentric distance (see lower black dashed
curve in top panel of Figure \ref{fig9}). During this late stage, the
inclination remains nearly constant and close the value excited by close
encounters between planets.

Interestingly, the pericenter argument of planet $m_1$ shows no signs of being
affected by the Lidov-Kozai cycles during most of the evolution, including the
phase when its inclination increased. A detailed analysis of this system shows
that the eccentricity/inclination excitation occurred in this system mainly due
to the effect of close encounters between planets, rather than to the
Lidov-Kozai (or secular) effects.

A second case of orbit evolution is shown in Figure \ref{fig10}, and
corresponds to Pop-I hot Jupiter on a prograde orbit. In this case, the
planet that became hot Jupiter was originally the outermost body in the system
(denoted by $m_2$).

The orbital evolution of all planets is characterized by a prolonged phase of
chaotic evolution, when planets move on crossing orbits and interact with each
other. During the later stages, the inclinations of $m_2$ and $m_3$ are
periodically excited to values close to $50^\circ$, returning to lower values
each time the eccentricity suffers a jump. This anti-correlation of $e$ and $I$
is characteristic of the Lidov-Kozai resonance, but is also observed in the
circulation domain of $\omega$ in the secular three-body problem.

The evolution of $\omega$ shows some evidence of temporary capture in the
Lidov-Kozai resonance (when either $\omega_3 = 90^\circ$ or $\omega_3 =
270^\circ$; see the bottom panel in Fig. 10), but overall this seems to have
only a minor cumulative effect on the final inclination. Pericenter argument
rapidly precesses during the very late stage (after $t\sim1.4\times10^5$ yr)
due to the combined effects of tides, relativity and stellar oblateness.

Finally, Figure \ref{fig11} shows an example of a system that produced the
Pop-II hot Jupiter.  Here the outer planet ($m_3$) is ejected from the system
at $t \simeq 10^5$ years. The system evolves more regularly after this time. 
Planet $m_1$ becomes trapped in the Lidov-Kozai resonance, and
remains in the resonance until the semimajor axis drops to very small
values. The orbital inclination shows no significant long-term evolution in the
Lidov-Kozai resonance remaining close to the value attained immediately after
the escape of $m_3$.

A similar analysis of a large sample of our simulations indicates an important
difference between the runs resulting in Pop-I planets, and those generating
Pop-II planets. In the first case, due to the fact that all planets remain in
the system for the duration of the run, the dynamical evolution shows a
prolonged phase of strong interaction between planets, leading to the
excitation of their eccentricities and inclinations. Since planetary orbits
suffer very strong perturbations, hot Jupiter can form only if the orbital
pericenter drops to very small values, where it can be tidally
trapped. Consequently, practically all hot Jupiters that formed in these systems
belong to Pop I. Also, the system's strong chaoticity produces a larger spread
in the inclination distribution, including numerous retrograde orbits in Pop I.

Another consequence of strong interaction in these systems is that the
Lidov-Kozai resonance can be sustained only for short intervals of time, and
does not seem to be important for the formation of hot Jupiters (retrograde or
not). Indeed we found no clear examples of the standard Kozai migration (Wu \&
Murray 2003).

It seems instead that the high eccentricities and inclinations are attained due
to effects of secular resonances between pairs of planets. We thus
agree with Wu \& Lithwick (2011) that the dynamical process responsible for the
formation of hot Jupiters is not simply the Kozai resonance, but rather a sort
of secular chaos in the regime of high eccentricities \& inclinations.
The exact relative importance of the secular chaos and effect of close
encounters between planets, however, has yet to be established. The close
encounters can clearly generate the AMD needed for the secular chaos to
operate, thus explaining the initial orbits used by Wu \& Lithwick (2011), or
can even be more central to the formation of hot Jupiter, by continuously
pumping AMD into the system until $e \sim 1$.

The orbit evolution of Pop-II Jupiters is qualitatively different from
that of Pop-I Jupiters, but the mechanisms that operate to produce large
eccentricities/inclinations are the same.  After the 3rd planet's ejection in
the systems that produce Pop II, the outer perturber generally has a large
semimajor axis. This causes smaller perturbations on and slower evolution of
the inner planet orbit, and permits tidal capture at larger pericentric
distance. Since the evolution is less chaotic and close encounters less
frequent, the excitation of the inclination is modest.  The retrograde orbits
are therefore more rare in Pop II than in Pop I.

We tested how the results discussed above depend on the tidal model. For
example, we switched of the modifications of the equilibrium tidal model
described in Section 2 and/or used different values of $Q'_p$. We then
conducted a set of simulations starting from the same initial conditions as
above. Although the results of individual runs were sensitive the assumed tidal
model, which is expected from the stochasticity of planet scattering, the
overall statistical results were very similar to those discussed above. This
shows the origin of hot Jupiters is insensitive of details of the tidal
interaction.

\begin{figure}[tb!]
\includegraphics[width=0.45\textwidth,clip=true]{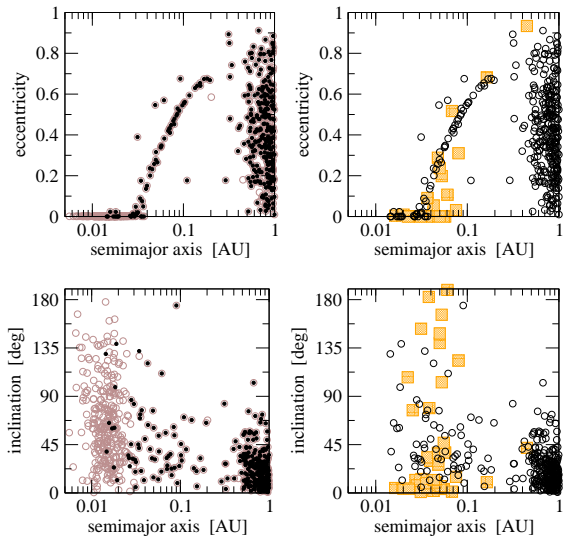}
\caption{Left: Orbit distribution of hot Jupiters that we obtained starting
  from the 3-planet systems.  Gray open circles show the transient population
  of orbits at $10^8$ yr, or the moment when the orbit of hot Jupiter became
  circularized by tides (defined as $e<0.01$). Black circles show orbits of planets 
  that survived the tidal decay for $T=10^9$ yrs, regardless of their
  eccentricity. Right: Comparison between the synthetic population (black
  circles) and known hot Jupiters with measured $|\lambda|$ (orange squares).}
\label{fig12}
\end{figure}

\subsection{Comparison with observed hot Jupiters}

To compare our results with observations, we should ideally follow each
planetary system for the estimated age of its host star. This is not practical,
however, because the analysis of thousands of synthetic systems and dozens of
different times would be complicated. We opted for a simplified comparison
instead.  First we continued the orbits of hot Jupiters, using our
semi-analytical method, to $1$ Gyr, which is a sort of the average age of known
hot-Jupiter's parent stars (e.g. Triaud 2011). Unlike in Sections 3.4 and 3.5.,
however, the hot Jupiters that reached the Roche radius of the star before $1$
Gyr were removed.\footnote{Recall that previously the orbital evolution was stopped when
the eccentricity reached $e = 0.01$, independently of the timescale when this
occurred.} The final distribution that we obtained here should therefore be
characteristic of 'aged' planetary systems.

Results are shown in the left-hand plots of Figure \ref{fig12}, where gray open
circles reproduce the data shown in Figure \ref{fig7}, while the black filled
circles show only those planets that survive at $1$ Gyr. Notably, most hot
Jupiters in Pop I disappear as they evolve by tides and are engulfed by the
star. Conversely, the orbit distribution of Pop-II Jupiters does not change
much. While Population-I hot Jupiters should therefore be expected only around
young stars, the Population II is more permanent.

The right-hand plots in Figure \ref{fig12} show the orbit distribution of
planets at $1$ Gyr (open black circles), while the orange squares show that of
the known hot Jupiters. The distribution of semimajor axes and eccentricities
of hot Jupiters produced in our simulations closely matches observations. There
is a clear trend, both in our results and observations, that hot Jupiters with
larger semimajor axes tend to have larger eccentricities. We discussed this in
Section 2.4.

In our simulations, the correlation between final semimajor axes and
eccentricities appear as a well defined curve, while the real hot Jupiters seem
to be more spread on both sides. This spread may be produced by uncertainties in 
the estimation of planetary eccentricities, or by a spread of $Q'_p$ values 
of real exoplanets. Also, with the Population I now completely removed, our 
simulations show no hot Jupiters with $a<0.015$~AU, exactly as observed.

For real planets we assumed that the measured $\lambda$ can be interpreted as
the orbital tilt, and plot it together with the orbital inclinations of hot
Jupiters obtained in our simulations. Note $\lambda$ is the {\it projected}
misalignment angle; a small value of $\lambda$ therefore does not guarantee that
star's spin vector and planet's orbit normal are actually aligned.\footnote{To
  compare simulations and observations more precisely, we would need to
  generate orbit normal vectors having an inclination distribution that our
  model predicts, and random orientation of nodes. This distribution should
  then be projected to the observer plane, and compared with observed
  distributions of $\lambda$ (e.g., Fabrycky \& Winn 2009).}

The distributions of both $I$ and $|\lambda|$ are broad, covering the whole range
from 0 to 180$^\circ$, and roughly similar (Figure \ref{fig12}, bottom right
panel), which is encouraging.  There are several, potentially important
differences between these two distributions as well. For example, the
inclination distribution of surviving hot Jupiters that we obtain in our
simulations shows the vast majority of prograde orbits, while the measured
values of $\lambda$ indicate that a relatively large fraction ($\sim$20\%) of
hot Jupiters are retrograde. Interestingly, as we discuss in the following
section, this problem may be resolved if the planetary systems had more than
three planets initially (i.e., before scattering).

\section{Four planet systems}

Here we consider the planetary systems with four planets initially. We used the
method described in Section 3.3 to set up the initial orbits and masses of four
planets. Specifically, the masses of planets were chosen according to
Eq. (\ref{eq16}), with an additional restraint that the lightest planet has
mass $>0.4$ M$_{\rm Jup}$. In total, 2166 planetary systems were followed. Of
these, 498 produced hot Jupiters (as defined by final $a < 0.1$ AU), including
both Pop I and Pop II.  This is $23 \%$, a fraction twice as high as the one
obtained with the three planet systems.

\begin{figure}[tb!]
\includegraphics[width=0.45\textwidth,clip=true]{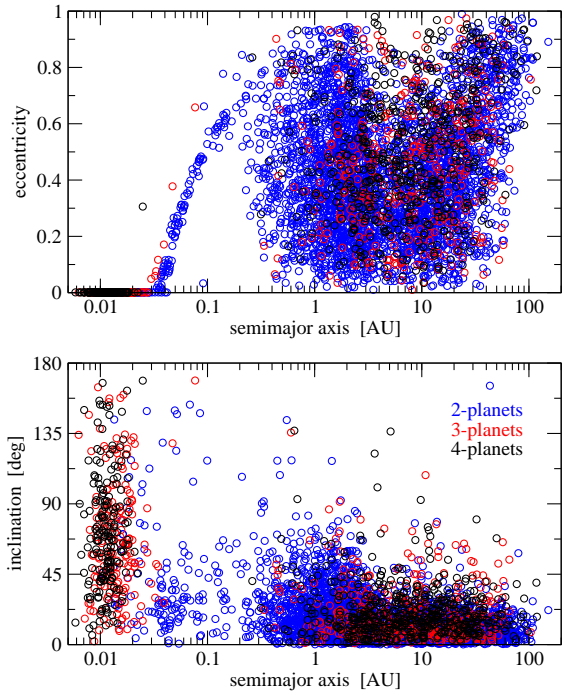}
\caption{Eccentricity (top) and inclination (bottom) distribution of 
 planetary orbits resulting from our 4-planet simulations. Color indicates the 
number of planets surviving in each system (i.e., blue denotes systems in which two of
the initial four planets were ejected, and two survived). Here we stopped 
the simulation if hot Jupiter formed and acquired a nearly circular orbit 
(defined as $e<0.01$).}
\label{fig13}
\end{figure}

The eccentricity and inclination distribution of exoplanets is shown in Figure
\ref{fig13}, where we have already separated cases according to the final
number of planets in each system. These results were obtained by numerically 
integrating the systems for $T=10^8$ yr, and extending these simulations to 
$T=10^9$ yr using our semi-analytical model. We show the final orbits of all 
planets at $T=10^9$ yr or, if hot Jupiter reached $e<0.01$, we show the orbits
at that time instant.

\begin{figure}[tb!]
\includegraphics[width=0.45\textwidth,clip=true]{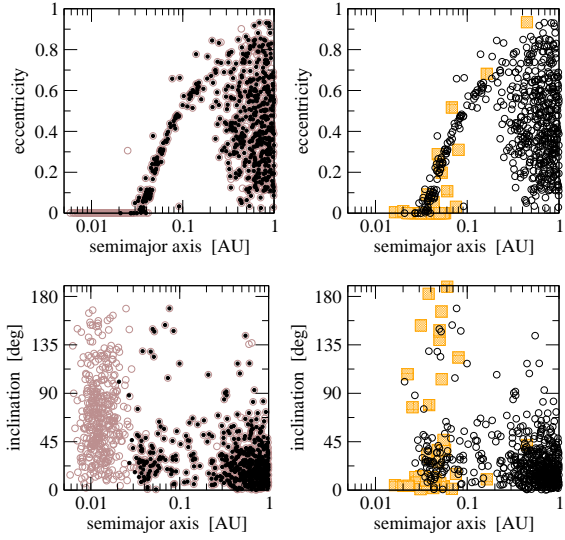}
\caption{Same as Figure \ref{fig12}, but for our four-planet simulations
  described in Section 4.}
\label{fig14}
\end{figure}

As in the 3-planet simulations discussed above, we find two distinct
populations of hot Jupiters: Population I with $a \le 0.03$ AU, and Population
II with $a > 0.03$ AU. As before, the Pop-II planets form in the systems that
end up with two planets (two planets are ejected in this case), while planet
ejection was less common in those systems that produced Pop I. Interestingly,
however, the inclination distribution of Pop-II hot Jupiters is now broader and
includes a larger fraction of retrograde orbits than in the 3-planet case.

A detailed analysis of individual 4-planet simulations shows similar mechanisms
at work as for the 3-planet systems (see Section 3.5). The orbit evolutions are
complex, show a dominant effect of planetary encounters and slow secular
interactions that are typically not related to the Lidov-Kozai resonance.

Finally, the left-hand panels in Figure \ref{fig14} show the final orbits of
hot Jupiters at $1$~Gyr.  This result confirms those obtained with the 3-planet
systems. The Pop-I hot Jupiters do not generally survive, while the orbits in
Pop II do not change much at late stages. Both the semimajor axis and
eccentricity distribution of final orbits show a good match to
observations. The match to the observed 3-day-period pile-up of hot Jupiters is
particularly good.

Unlike in the three planet case, the inclination distribution obtained with
four initial planets shows a relatively large fraction of retrograde orbits
($\sim$10\%). Also, some of the retrograde orbits now have $I>150^\circ$, as
required to match several known systems with $|\lambda|\gtrsim150^\circ$. For
comparison, observations indicate that $\sim20$\% of hot Jupiters have
$|\lambda|>90^\circ$. It is not clear at this point whether the difference
between simulated 10\% and observed 20\% is significant, mainly because the
observational statistics is still pretty low.

\section{Conclusions}

Here we reported a series of simulations in which we followed the orbital and spin
evolution of planetary systems starting with three and four planets. The
planets were initially placed in resonant orbits as expected from their
formation and migration in the protoplanetary gas disk. The instability was
triggered in each of these systems by breaking the resonant locks. Soon after
the onset of instability, planets scatter each other and typically obtain large
orbital eccentricities and inclinations. In addition to the gravitational
interactions between planets, we also included the effects of relativity,
stellar oblateness and tides, which are mainly important for orbits with low
pericenter distance.  We modified the standard equilibrium model to mimic the
effects of dynamical tides for $e\sim1$.

We found that scattering and subsequent slow secular interaction between
planets generate a number of planets in quasi-parabolic orbits that tidally
evolve. A fraction of these planets can survive over Gyr timescales. The
surviving population provides a good match to observations, including the
3-day-period pile-up of hot Jupiters, nearly circular orbits of hot Jupiter for
$\lesssim 0.03$ AU, correlation between $a$ and $e$ for orbits with larger $a$,
etc.

Contrary to previous works, we found the the Kozai resonance is {\it not} the
dominant evolution path leading to the formation of hot Jupiters. The vast
majority of hot Jupiters in our simulations acquire small pericentric distances
(and in some cases retrograde orbits) by being scattered by other planets
and/or during the subsequent phase of slow secular evolution (typically
unrelated to the Kozai resonance).

We find that $\approx10 \%$ of planetary systems starting with three planets
produce hot Jupiters, while this ratio increases to $\approx23$\% if four
planets are considered. However, most of these are eliminated by subsequent
tidal decay for timescales of the order of $1$ Gyr. The proportion of surviving
hot planets drop to $\approx2$\% for 3-planet systems and $\approx5$\%
for our 4-planet runs. 

In both cases, we find that hot Jupiters can be divided
into two populations. The transient Population-I hot Jupiters typically form in
the systems where no (in the 3 planet case) or up to one planet (in the 4 planet
case) is ejected, have $a<0.03$ AU, and a very broad inclination distribution, 
including a large fraction of retrograde orbits. These planets
continue to evolve tidally and generally do not survive $1$ Gyr.  They are
expected to be present only around younger stars.

The Population-II hot Jupiters form in the systems where one (in the
3 planet case) or two planets are ejected (in the 4 planet case), have $a>0.03$
AU, and a smaller fraction of retrograde orbits. The Pop-II hot Jupiters are
far enough from the central star to have only minimal tidal evolution, and
generally survive on Gyr-long time scales. They provide the best match to the
observed population of hot Jupiters.

We found that the initial systems with more than three planets tend to produce
hot Jupiters more often than the systems with three initial planets, and the
orbits of hot Jupiters born in the former systems tend to have broader
inclination distribution, including about 10\% of retrograde orbits. Both these
characteristics appear to provide a better match to observations (although much
work remains to be done). This may suggest that planetary systems emerging from
the protoplanetary disks, or at least the ones that seed hot Jupiters, are
initially unexpectedly rich in the number of planets.

\acknowledgements {\bf Acknowledgments:} This work has been supported by NSF
AAG. C.B. would like to express his gratitude to the Southwest Research
Institute and all its staff for invaluable help during the development of this
work.


\begin{thebibliography}{}

\bibitem[]{bea11}
Benitez-Llambay, P., Masset, F., Beaug\'e, C. 2011. A\&A, 528, A2.

\bibitem[]{beu05}
Beutler, G. 2005. ``Methods of Celestial Mechanics'', Springer-Verlag, Germany.

\bibitem[]{cha99}
Chambers, J.E. 1999. MNRAS, 304, 793.

\bibitem[]{cea08}
Chatterjee, S., Ford, E.B., Matsumura, S., Rasio, F.A. 2008. ApJ, 686, 580.

\bibitem[]{d79}
Darwin, G.H. 1879. The Observatory, 3, 79.

\bibitem[]{dea11}
Deleuil, M. et al. 2011. A\&A, submitted (astroph-1109.3203).

\bibitem[]{fw09}
Fabrycky, D.C., Winn, J.N. 2009. ApJ, 696, 1230. 

\bibitem[]{g93}
Gladman, B., 1993. Icarus, 106, 247.

\bibitem[]{h81}
Hut, P. 1981. A\&A, 99, 126.

\bibitem[]{ip04}
Ivanov, P.B., Papaloizou, J.C.B. 2004. MNRAS, 347, 437.

\bibitem[]{ip07}
Ivanov, P.B., Papaloizou, J.C.B. 2007. MNRAS, 376, 682.

\bibitem[]{ip11}
Ivanov, P.B., Papaloizou, J.C.B. 2011. CeMDA, 111, 51.

\bibitem[]{jea08}
Jackson, B., Greenberg, R., Barnes, R. 2008. ApJ, 678, 1396.

\bibitem[]{jt08}
Juric, M., Tremaine, S. 2008. ApJ, 686, 603.

\bibitem[]{k03}
Kley, W. 2003. CeMDA, 87, 85.

\bibitem[]{k62}
Kozai, Y. 1962. AJ, 67, 591.

\bibitem[]{l97}
Lai, D. 1997. ApJ, 490, 847.

\bibitem[]{lea11}
Lai, D., Foucart, F., Lin, D.N.C. 2011. MNRAS, 412, 2790.

\bibitem[]{lea09}
Lainey, V., Arlot, J,-E, Karatekin, \"O, Van Hoolst, T. 2009. 
Nature, 459, 957.

\bibitem[]{lr91}
Laskar, J., Robutel, P. 1995. CeMDA, 62, 193.

\bibitem[]{ld00}
Levison, H.F., Duncan, M.J. (2000). ApJ, 120, 2117.

\bibitem[]{lt11a}
Libert, A.-S., Tsiganis, K. 2011a. MNRAS, 412, 2353.

\bibitem[]{lt11b}
Libert, A.-S., Tsiganis, K. 2011b. CeMDA 111, 201.

\bibitem[]{l61}
Lidov, M.L. 1961. Isk. Sput. Zemli, 8, 119.

\bibitem[]{mb82}
Marchal, C., Bozis, G. 1982. CeMDA, 26, 311.

\bibitem[]{m07}
Mardling, R.A. 2007. MNRAS, 382, 1768.

\bibitem[]{mw02}
Marzari, F., Weidenschilling. S.J. 2002. Icarus, 156, 570.

\bibitem[]{m00}
Masset, F. A\&ASS, 141, 165.

\bibitem[]{mcea10} 
McArthur, B.E., Benedict, G.F., Barnes, R., Martioli, E.,
Korzennik, S., Nelan, E., Butler, R.P. 2010. ApJ, 715, 1203.

\bibitem[]{mt09}
Moeckel, N., Throop, H.B. 2009. ApJ, 707, 268.

\bibitem[]{m79}
Mignard, F. 1979. Moon and Planets, 20, 301. 
 
\bibitem[]{m80}
Mignard, F. 1980. Moon and Planets, 23, 185. 
 
\bibitem[]{mea07}
Morbidelli, A., Tsiganis. K., Crida, A., Levison, H.F., Gomes, R. 2007. AJ, 134, 1790.

\bibitem[]{mea11}
Moutou, C., et al., 2011. A\&A, 533, A113.

\bibitem[]{nea08}
Nagasawa, M., Ida, S., Bessho, T. 2008. ApJ, 678, 498.

\bibitem[]{nea11}
Naoz, S., Farr, W.M., Lithwick, Y., Rasio, F.A., Teyssandier, J. 2011. Nature, 473, 187.

\bibitem[]{pn08}
Pierens, A., Nelson, R.P. 2008. A\&A, 482, 333.

\bibitem[]{s10}
Schlaufman, K.C. 2010. ApJ, 719, 602.

\bibitem[Skumanich(1972)]{1972ApJ...171..565S} Skumanich, A.\ 1972, \apj, 
171, 565 

\bibitem[]{sea01}
Snellgrove, M.D., Papaloizou, J.C.B., Nelson, R.P. 2001. A\&A, 374, 1092.

\bibitem[]{tea08}
Thommes, E.W., Bryden, G., Wu, Y., Rasio, F.A. 2008. ApJ, 675, 1538.

\bibitem[]{t11}
Triaud, A.H.M.J. 2011. A\&A, submitted (astroph 1109.5813v1).

\bibitem[]{tb08}
Throop, H.B., Bally, J. 2008. AJ, 135, 2380.

\bibitem[]{wm03}
Wu, Y., Murray, N. 2003. ApJ, 589, 605.

\bibitem[]{wl11}
Wu, Y., Lithwick, Y. 2011. ApJ, 735, 109.

\end{thebibliography}
\end{document}